\documentclass{jetpl}
\usepackage{graphicx}
\usepackage{ulem}
\usepackage[dvipsnames]{xcolor}

\usepackage{datetime2}

\graphicspath{{../figures}}
\twocolumn

\lat

\DeclareMathOperator{\Tr}{Tr}

\title{M{\o}lmer-S{\o}rensen gates in trapped-ions chains in the presence of correlated noise}

\rtitle{M{\o}lmer-S{\o}rensen gates \ldots}

\sodtitle{M{\o}lmer-S{\o}rensen gates in trapped-ions chains in the presence of correlated noise}

\author{
D.\,V.\,Donchenko$^{+*}$,
E.\,A.\,Anikin$^+$\/\thanks{e-mail: evgenii.anikin@skoltech.ru},
O. Lakhmanskaya$^+$,
K. Lakhmanskiy$^+$}

\rauthor{Donchenko D.V., Anikin E.V., Lakhmanskaya O. Lakhmanskiy. K}

\sodauthor{Donchenko, Anikin, Lakhmanskaya, Lakhmanskiy}

\address{${}^+$Russian Quantum Center,
117940 Moscow, Russia\\~\\
$^*$National Research Nuclear University MEPhI, Moscow 115409, Russia}

\dates{\today}{*}

\abstract{
  We analyze the impact of correlated laser frequency noise on 
  M{\o}lmer-S{\o}rensen gates in qubit registers based on 
  trapped-ion chains. Using perturbation theory, we 
  calculate gate fidelities in the presence of noise with arbitrary 
  power spectral density for different chain lengths 
  and ion positions in the chain. With our approach, we account 
  for simultaneous excitation of multiple phonon modes during gate operation.
  We find out that the impact of medium-frequency laser noise depends 
  considerably on the positions of the ions in the chain. In contrast,
  low-frequency noise has similar effect for different chain lengths 
  and ion positions.
}

\begin{document}

\maketitle

\section{Introduction}
Practical applications of quantum computing require the
capability to implement high-fidelity quantum gates within
large qubit registers.  However, the performance of
up-to-date devices is limited by imperfect control and
noises.  This makes noise analysis the key component of the
design of quantum computing devices.  However, such an
analysis becomes increasingly difficult for large qubit
registers, which are many-body interacting systems requiring
exponential amount of computational resources for
simulation.

One of the most successful candidate systems for the
implementation of a quantum computer are cold trapped-ion
chains \cite{SchmidtKaler2003, MonzKimVillar2009,
Bermudez2017, Bruzewicz2019, Pogorelov2021}.  
In cold trapped ions, one of the most frequently 
used entangling operation os the 
M{\o}lmer-S{\o}rensen (MS) gate \cite{Soerensen2000}
which is implemented by a bichromatic
laser pulse symmetrically detuned from the ion qubit
transition. The pulse creates a qubit-state-dependent
displacement of the collective phonon modes of the ion chain
\cite{Haljan2005}.  The resulting action of the pulse on the
internal qubit states reduces to the $R_\mathrm{XX}(\theta)$
gate \cite{NielsenChuang2010} providing that the
displacement of all phonon modes vanish at the end of the
pulse. In long chains, the latter condition requires special
design of laser pulse shapes \cite{Zhu2006a, Choi2014,
Leung2018}. For up-to-date MS gate
implementations, the best values of the gate fidelity can
reach $0.99{-}0.999$. However, even larger fidelities
($>0.9999$) are required for fault-tolerant quantum
computation \cite{Bermudez2017}.

To achieve larger MS gate fidelities, it is
necessary to account for the contributions of various
unwanted effects and to establish bounds on the acceptable
noise levels.
However, modeling gate dynamics in long chains beyond an
exactly solvable spin-dependent force (SDF) Hamiltonian
presents a computational challenge.
There exists a range of the 
additional dynamical contributions to the laser-ion
Hamiltonian, such as carrier transition,
out-of-Lamb-Dicke terms, nonlinearity of the Coulomb
potential, micromotion, and other \cite{Wu2018, Saner2023,
Bazavan2023, Bluemel2024, Anikin2025, OrozcoRuiz2025}.
Another group of corrections to the SDF model arise from 
the imperfections of the control parameters and noises, 
such as the laser field, magnetic 
field, and trapping potential fluctuations \cite{Wu2018,
Ballance2017}. These contributions make the time-dependent 
Schr{\o}dinger equation (TDSE) for qubit-phonon system not solvable 
exactly. Furthermore, the numerical TDSE solution is not
possible for long ion chains: to account for the excitation of 
multiple phonon modes during gate operation, the
exponentially large Hilbert space is required.


In this manuscript, we focus on the effect
of the laser frequency noise on the MS gate implementation
in long ion chains. It is known as one of the dominant
sources of the MS gate error \cite{Ballance2017}. Also,
the noise spectrum of typical laser sources used for
trapped-ion quantum state manipulation can considerably
vary between setups and contain contributions on the
timescale of the inverse gate duration
\cite{KirchmairDissertation, ManchaoZhang2021}. Because of
that, a considerable effect of the correlated laser noise
on the MS gate could be expected. To take it into account,
we develop a perturbative approach to analyze the gate
dynamics in the presence of noise with
arbitrary power spectral density.  We show
that for the frequency noise with power spectral density
$S(\omega)$, the gate infidelity can be expressed as
$\frac{1}{2\pi}\int_0^{+\infty} d\omega\,
S(\omega)\tilde{I}(\omega)$, where the \textit{sensitivity
function} $\tilde{I}(\omega)$ does not depend on the noise
properties and can be inferred from the ideal gate dynamics.

We calculate the sensitivity functions for MS gates for gate
and chain parameters typical for up-to-date experiments. In
particular, we consider ion chains of different lengths and
different ion positions in the chain. We find that the
contribution of the low frequency noise (lower than the
inverse gate time) to the MS gate infidelity is almost
independent on the considered parameters. In contrast, the
conribution of the medium frequency noise (of the order of
the inverse gate time) depends considerably on the ion
positions in the chain. In particular, we find that the MS
gate between distant ions is more sensitive to the medium
frequency noise than for neighbor ions.

Our method establishes a rigorous framework for
characterization of noise properties of trapped-ion quantum
computers. The developed perturbative approach allows
predicting gate infidelities in long trapped-ion chain
subject to noisy laser field. Our analysis reveals
nontrivial dependencies of the gate sensitivities to noise
on the chain length and ion positions within the chain.

\section{The model}
\begin{figure}
  \begin{minipage}{0.34\linewidth}
    \includegraphics[width=\linewidth]{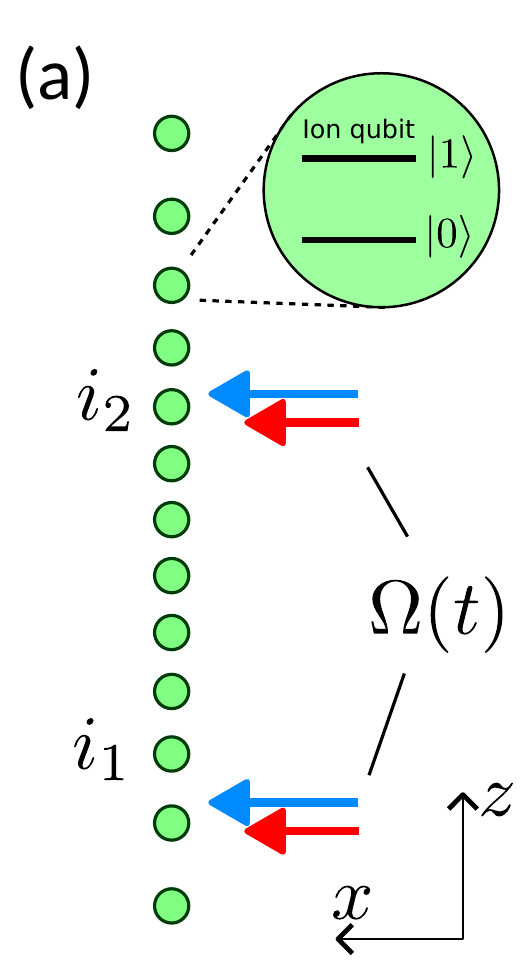}
  \end{minipage}
  \begin{minipage}{0.65\linewidth}
    \includegraphics[width=\linewidth]{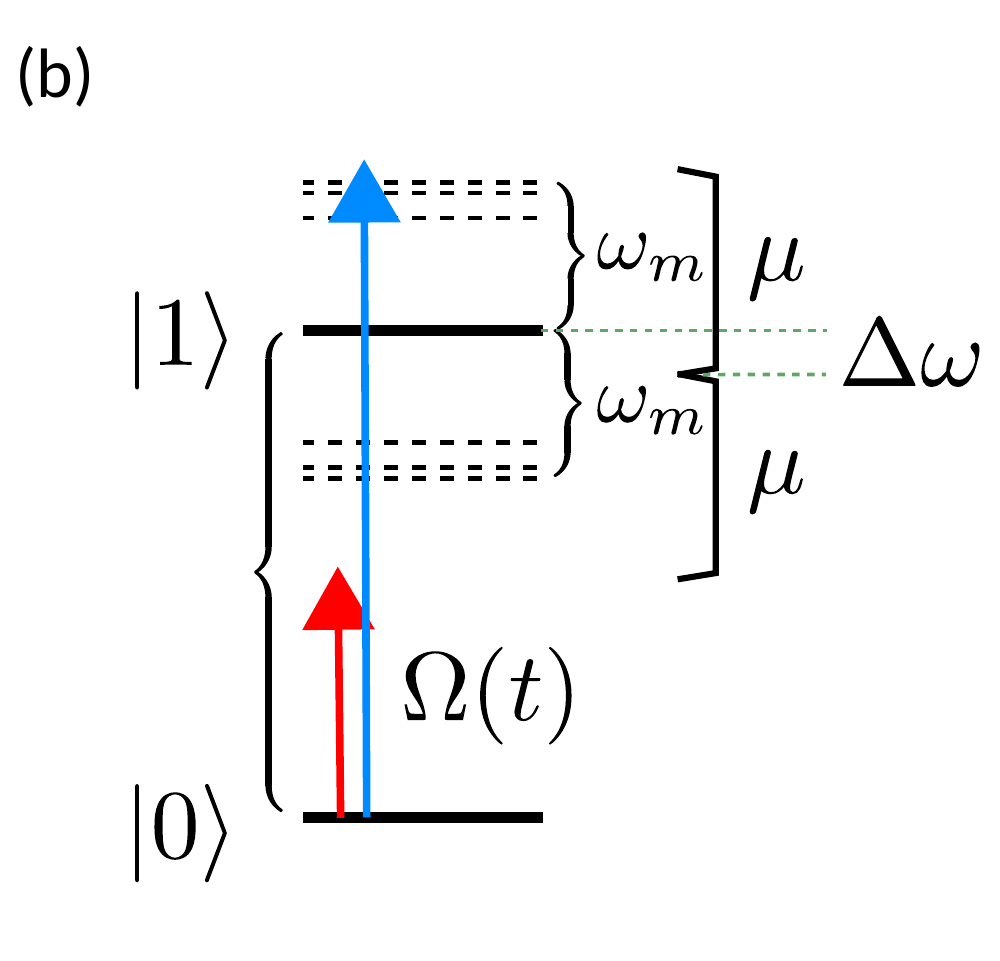}
  \end{minipage}
  \caption{Figure \thefigure.  (a) The ion chain in the Paul trap. The ions $i_1$, $i_2$ are illuminated
  by bichromatic beams perpendicular to the ion chain. (b) The bichromatic beam detunings
  $\pm\mu$ are set close to the blue and red motional sidebands $\pm\omega_m$ of the chain.
  The laser frequency fluctuation shifts the frequencies of both components by $\Delta\omega$.}
  \label{fig:setup}
\end{figure}

We consider the MS gate implementation in a linear trapped-ion chain. Two target ions
$i_1$, $i_2$ are illuminated by bichromatic beams perpendicular to the ion chain (see Fig.~\ref{fig:setup}).
For the ideal gate operation, the beam components should be symmetrically detuned from the 
qubit transition. Under these conditions, the beams create qubit state-dependent forces 
exciting transverse ions motion.  The Hamiltonian for the interaction of the target ions
with $n$ radial phonon modes is as follows \cite{Choi2014, Zhu2006, Monroe2021}:
\begin{equation}
  \hat H_0 = \sum_{i\in{i_1,i_2}}\Omega(t) \cos(\mu t) k\hat{x}_i(t)
  \sigma_x^{(i)}
  \label{eq:sdf_ham}
\end{equation}
\begin{equation}
  k\hat{x}_i(t) = \sum_{m}\eta_{im}(a_m e^{-i\omega_m t} + a_m^\dagger e^{i\omega_m t }),
\end{equation}
where $\Omega(t)$ is the bichromatic beam amplitude, $\mu$ is the detuning of the bichromatic beam
from the carrier transition, $k$ is the laser wavevector, $\eta_{im}$ are the Lamb-Dicke parameters,
$\hat{a}_m^\dagger$ and $\hat{a}_m$ are the creation and annihilation operators of the phonon modes,
and $\omega_m$ are the frequencies of the phonon modes. We discuss the calculation of these parameters
in Appendix~\ref{appendix:phonon_modes}.

The Hamiltonian \eqref{eq:sdf_ham} is a spin-dependent force Hamiltonian. It is exactly solvable, 
and the evolution operator takes form \cite{Monroe2021}
\begin{multline}
  \hat{U}_0(t_1, t_2) = 
    \exp\left(-\frac{i}{2}\sum_{i,j} \chi_{i,j}(t_2, t_1) \sigma_{x}^i \sigma_{x}^j\right)\\
    \prod_m D_m\left(\sum_i\sigma^i_x\alpha_{im}(t_2, t_1)\right),
    \label{eq:ms_evo_op}
  \end{multline}
where $D_m(\alpha)$ is the displacements operator for phonon mode $m$, and
\begin{equation}
  \alpha_{im}(t_2, t_1) = -i\int_{t_1}^{t_2} f_{im}(t') dt',
  \label{eq:alpha}
\end{equation}
\begin{equation}
  f_{im} = \eta_{im} e^{i\omega_m t}\Omega(t) \cos{(\mu t + \psi)},
\end{equation}
\begin{equation}
\chi_{ij}(t_2, t_1) = 2\Re\int_{t_1}^{t_2} \alpha_{im} f_{jm}^* dt'.
  \label{eq:chi}
\end{equation}
The pulse shape $\Omega(t)$ should be chosen to satisfy
the condition $\chi_{i_1i_2} = \pi/4$ while minimizing the
error caused by the imperfectly closed 
phonon mode phase trajectories \cite{Landsman2019}:
\begin{equation}
  1-F_0 = \sum  |\alpha_{im}|^2.
  \label{eq:unclosed_traj_err}
\end{equation}
The Eq.~\eqref{eq:unclosed_traj_err} represents the error for the initial qubit state $|00\rangle$.

In this manuscript, we consider the amplitude-modulated pulse shapes defined 
as piecewise cubic polynomials (splines) \cite{Anikin2025}.

\section{MS gate in the presence of laser frequency noise}
The laser frequency noise $\Delta\omega(t)$ can be modeled as a time-dependent
qubit frequency shift. Therefore, the laser-ion Hamiltonian reads
\begin{equation}
  \hat{H} = \hat{H}_0 + \hat{V},
  \label{eq:ham_w_freq_noise}
\end{equation}
where
\begin{equation}
  \hat{V} = \Delta \omega(t)\hat{S}_z,
\end{equation}
and $\hat{S}_z = \frac{1}{2}(\sigma_z^{(i_1)} + \sigma_z^{(i_2)})$. Because of the presence of the non-commuting 
$\sigma_x$ and $\sigma_z$ operators, the 
Hamiltonian \eqref{eq:ham_w_freq_noise} is not exactly solvable. Below, 
we utilize perturbation theory in $\hat{V}$ to find its effect on 
MS gate dynamics.

The evolution operator in the presence of the 
frequency fluctuation $\hat{V}$ reads
\begin{equation}
  \hat{U} = \hat{U}_0(1 - i\hat{T}).
\end{equation}
Here $\hat{U}_0$ is the ideal MS gate operator, and $\hat{T}$ reads
\begin{equation}
  \hat{T}(t) = \int_{t_0}^{t} dt' \hat{V}_I(t'),
\end{equation}
where $\hat V_I(t) = \Delta\omega(t) U_0^\dagger(t, t_0) \hat{S}_z U_0(t, t_0)$ is the perturbation operator in 
the interaction picture. For the initial state $|\psi_0\rangle$, the infidelity of the resulting state can 
be calculated as
\begin{multline}
  1 - F = \operatorname{Tr}_{\text{ph}}
\left[\langle\psi_0|T^\dagger T |\psi_0\rangle - 
\langle\psi_0|T^\dagger|\psi_0\rangle\langle\psi_0| T |\psi_0\rangle
\right]\\
= 
\int_{}^{}dt\,dt'\Delta\omega_{L}\left( t \right)\Delta\omega_{L}\left( t' \right)I\left( t,t' \right)
  \label{eq:infidelity}
\end{multline}
where \(I\left( t,t' \right)\) reads
\begin{multline}
  I\left( t,t' \right) = \text{Tr}_{\text{ph}}
  \left[ \left\langle \psi_{0}\left| S_{z}^{I}\left( t \right)S_{z}^{I}\left( t' \right) \right|\psi_{0} \right\rangle \right.\\
  - \left. \left\langle \psi_{0}\left| S_{z}^{I}\left( t \right) \right|\psi_{0} \right\rangle\left\langle \psi_{0}\left| S_{z}^{I}\left( t' \right) \right|\psi_{0} \right\rangle \right]
  \label{eq:sensitivity_function_time}
\end{multline}
The trace over phonons in the definition of $I(t,t')$ can be calculated using the 
analytical form of the evolution operator \eqref{eq:ms_evo_op} for any number of phonon modes: more details are 
in the Appendix~\ref{appendix:trace_calculation}

For a random frequency noise $\Delta\omega(t)$ with a known correlation function $C(t,t')$, the gate infidelity can be 
found by averaging $\eqref{eq:infidelity}$.
Infidelity can also be conveniently expressed though the noise power spectral density $S(\omega)$:
\begin{equation}
  1-F = \int_0^{\infty} \frac{d\omega}{\pi} \tilde I(\omega) S(\omega),
  \label{eq:corr_noise_infidelity}
\end{equation}
where 
\begin{equation}\tilde{I}(\omega) = \int_{}^{}dt dt'
  \cos\left[ \omega(t-t')\right] I\left( t,t' \right).
  \label{eq:sensitivity_function_frequency}
\end{equation}
We call the function $\tilde{I}(\omega)$ the \textit{sensitivity function}
of the MS gate to the frequency noise. For a narrow-band noise 
at the frequency $\omega_0$ with
mean square deviation $\Delta\omega$, 
the power density spectrum can be taken as 
$S(\omega) = \pi\Delta\omega^2\delta(\omega-\omega_0), \,
\omega>0$. 
Therefore, the infidelity in this case is proportional
to $\tilde{I}({\omega_0})$:
\begin{equation}
  1-F = \Delta\omega^2\tilde{I}({\omega_0}).
  \label{eq:narrow_band_infidelity}
\end{equation}
The sensitivity function $\tilde{I}(\omega)$ does not depend on the 
noise properties. Therefore, for the given pulse $\Omega(t)$ implementing
the MS gate, the sensitivity function can be used to calculate 
the gate infidelity for noises with different power spectral densities.
However, $\tilde{I}(\omega)$ depends on the details of the ideal gate
implementation. It depends on the chain phonon mode spectrum, the ion
positions $i_1$ and $i_2$, and the laser pulse $\Omega(t)$ implementing 
the gate.

We calculate the sensitivity functions using Eqs.
\eqref{eq:sensitivity_function_frequency}, 
\eqref{eq:sensitivity_function_time} for MS gates 
with different parameters. For all cases, we take the
initial ion wavefunction as $|\psi_0\rangle = |00\rangle.$ We consider different  
chain lengths and ion positions in the chain. For each case, 
the pulse shape $\Omega(t)$ is found from a 5-parametric 
family of piecewise polynomials by minimization of the 
error \eqref{eq:unclosed_traj_err} 
caused by imperfectly closed phonon phase trajectories. 
For all considered cases, we choose the same 
gate time of $310 \mu\mathrm{s}$ and the detuning 
$\delta = (2\pi) 8.06 \mathrm{kHz}$ from the the trap secular 
frequency in $x$-direction.
For these parameters, our method for pulse shape calculation 
yields a simple bell-like shape. For all
considered cases, the contribution of imperfectly closed phase
trajectories into gate error does not exceed $10^{-4}$.

\begin{figure}
  \includegraphics[width=\linewidth]{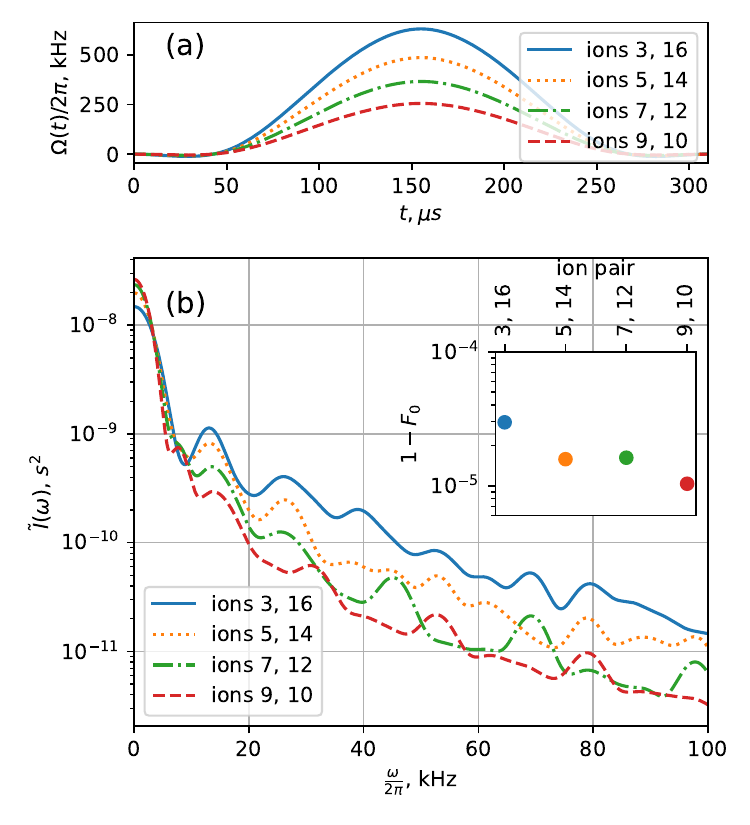}
  \caption{Figure \thefigure. (a) The pulse shapes $\Omega(t)$ for the implementation of the MS gate 
    between different ions in a chain of 20 ${}^{40}\mathrm{Ca}^+$ ions.  The trap axial and radial 
    frequencies are set to $\omega_z = (2\pi) \,147.2 \,\mathrm{kHz}$ and $\omega_x = (2\pi)\, 2.7 \,\mathrm{MHz}$, 
    and the bichromatic detuning is 
    $\mu = \omega_x + \delta$, $\delta = (2\pi)\, 8.06 \,\mathrm{kHz}$.
    (b) The MS gate sensitivity functions $\tilde{I}(\omega)$ 
    for the pulse shapes shown in (a).
  The inset: the contribution of imperfectly closed phase trajectories into error.}
  \label{fig:noise_sensitivities_diff_ions}
\end{figure}

In Fig.~\ref{fig:noise_sensitivities_diff_ions}, we show the 
sensitivity functions $I(\omega)$ for MS gates in a 20-ion chain
between different ion pairs. In all cases, the $\tilde I(\omega)$ have
a maximum of $\sim 10^{-8}\, \mathrm{s}^2$ at $\omega=0$. According 
to Eq.~\eqref{eq:narrow_band_infidelity}, this implies that 
slow frequency fluctuations by $\sim 1\,\mathrm{kHz}$ cause gate infidelities of 
order of percents. Also, the sensitivity functions decay with increasing 
$\omega$. Importantly, the decay is slower for distant ions, which indicates 
larger sensitivity of the MS gate between distant ions to noise at nonzero
frequency. For example, for $\omega \sim (2\pi)\, 25\mathrm{kHz}$, the sensitivity 
function for MS gate between ions $3$ and $16$ is almost by order of magnitude larger 
than between the neighbor ions $9$ and $10$.

\begin{figure}
  \includegraphics[width=\linewidth]{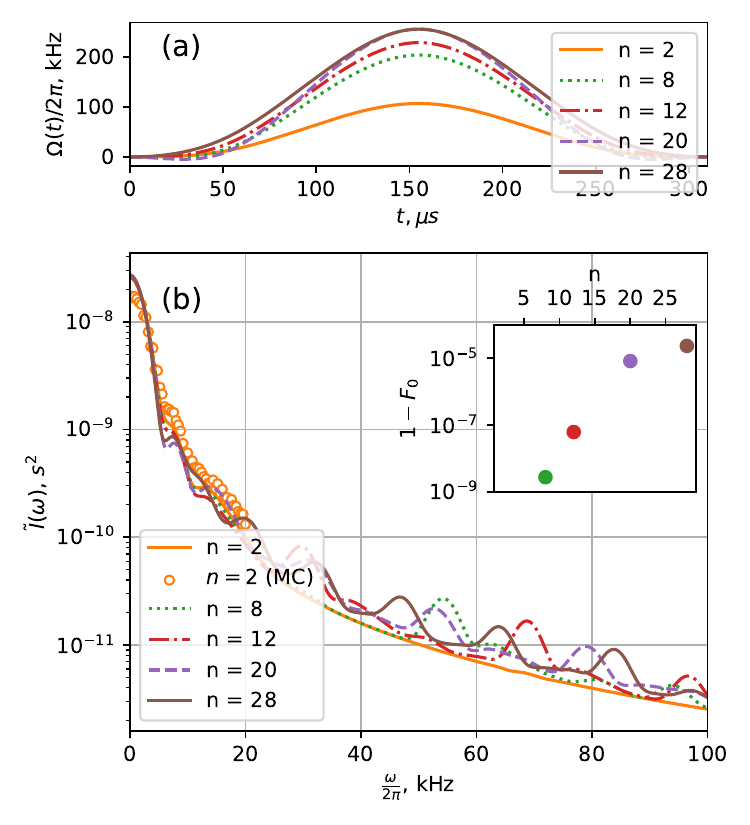}
  \caption{Figure \thefigure.\, (a) 
    The pulse shapes $\Omega(t)$ implementing MS gate
    between two central ions in ${}^{40}\mathrm{Ca}^+$ ion
    chains of different lengths.
    The trap radial frequency and the bichromatic detuning are the same as for Fig.~\ref{fig:noise_sensitivities_diff_ions}, 
    and the axial frequency is chosen differently for each $n$ to keep the smallest radial frequency equal to
    $(2\pi) 2.4 \mathrm{MHz}$.
    (b) The MS gate sensitivity functions $\tilde{I}(\omega)$ 
    for the pulse shapes shown in (a). The circles show the results of 
    the Monte-Carlo simulations for two ions.
    The inset: the contribution of imperfectly closed phase trajectories into error.}
  \label{fig:noise_sensitivities_central_ions}
\end{figure}

In Fig.~\ref{fig:noise_sensitivities_central_ions}, we show the 
sensitivity functions for MS gates in the chains
of different lengths for two ions in the middle of the chain. In this case,
$I(\omega)$ decays similarly for different chain length. However, for certain frequencies, 
the sensitivities in long chains exhibit peaks which are absent for shorter chains. Also, we compare
the analytical calculation of the sensitivity function with the Monte-Carlo simulations
of the Hamiltonian \eqref{eq:ham_w_freq_noise} for a 2-ion chain. We calculated 
the average MS gate infidelity over 200 generated samples of the frequency noise
with mean square deviation $\Delta\omega = (2\pi) 500\mathrm{Hz}$ and 
correlation time of $500\mu\mathrm{s}$. With the correlation time exceeding 
gate duration, the approximation of \eqref{eq:narrow_band_infidelity} can be used
to find the sensitivity function from the calculated values of the infidelity. The 
sensitivity function obtained from Monte-Carlo simulations is in agreement with 
the analytical calculations using Eq.~\eqref{eq:sensitivity_function_frequency} 
(see Fig.~\ref{fig:noise_sensitivities_central_ions}(b)).

\section{Conclusions}
We analyzed M{\o}lmer-S{\o}rensen gates in long trapped-ion chains in the presence 
of correlated laser frequency 
noise. With the help of the leading-order perturbation theory, we found expressions for MS gate infidelity for arbitrary noise 
power spectral density. Using these expressions, we analyzed the impact of noise on MS gates with different parameters,
in particular, different chain lengths and different ion positions in the chain. We found out that MS gates between distant ions 
in long chains are more sensitive to medium-frequency noise (tens of kilohertz) than neighbor ions. In contrast, the sensitivity
to the low-frequency noise depends weakly on the chain length and ion positions. Our results allow precise
examination of noise properties of the trapped-ion quantum computers and could be useful for the design of trapped-ion 
devices with minimal noise impact.

\section*{Acknowledgements}
The work was supported by Rosatom in the framework of the Roadmap for Quantum
computing (Contract No. 868/1759-D dated 3 October 2025).

\bibliography{references}

\appendix
\onecolumn
\section{Radial phonon modes of the ion chain}
\label{appendix:phonon_modes}
The interaction of the laser radiation with the ion chain phonon modes is defined by 
the matrix of the Lamb-Dicke parameters $\eta_{im}$ defined as 
\begin{equation}
  \eta_{im} = k\sqrt{\frac{\hbar}{2M\omega_{m}}}b_{im},
\end{equation}
where $k$ is the wavevector of the laser radiation, 
$M$ is the ion mass, $\omega_m$ are the frequencies of the phonon modes, 
and $b_{im}$ are the normal vectors. The frequencies and normal vectors
can be obtained from the Hamiltonian for ion motion in the trap \cite{James1998}
\begin{equation}
  H_\mathrm{motional} = \sum_{i} \frac{(\vec p_i)^2}{2M} + 
  U_\mathrm{trap}(\vec{r}_i) 
  + \sum_{i<j} \frac{e^2}{4\pi\epsilon_0 |\vec{r}_i - \vec{r}_j|},
\end{equation}
where $U_\mathrm{trap} = \frac{M}{2}(\omega_x x^2 + \omega_y y^2 + \omega_z z^2)$ 
is the trap potential. We assume that the trap
frequencies are chosen so that the ions form a linear configuration along 
the $z$ axis. To find the normal vectors and frequencies,
one should first find the ion equilibrium positions by minimization of 
the potential energy in $H_\mathrm{motional}$. Then, normal modes
can be found from the expansion around the equilibrium positions.

\begin{figure}
  \centering
  \includegraphics[width=0.6\linewidth]{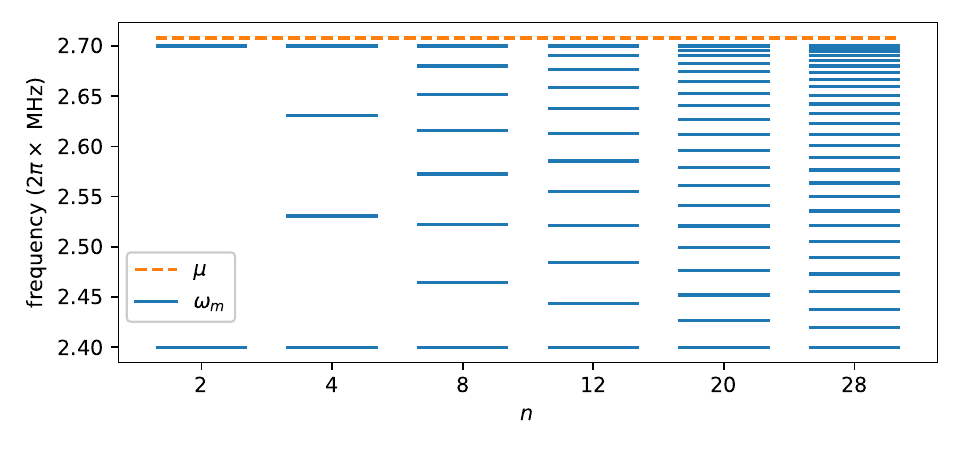}
  \caption{Figure~\thefigure. Blue solid lines show the radial phonon mode frequencies of the $n$-ion chain in
  the harmonic potential with the radial secular frequency $2.7 \mathrm{MHz}$. Orange dashed line
  shows the value of the bichromatic detuning $\mu$ chosen for MS gate implementation.}
  \label{fig:radial_normal_modes}
\end{figure}

In Fig.~\ref{fig:radial_normal_modes}, we show the calculated values of the 
normal mode frequencies along the $x$ axis. We used the 
trap secular frequency of $2.7 \mathrm{MHz}$ in $x$-direction. For each number
of ions, the axial frequency was adjusted so that the minimal radial phonon frequency
in $x$-direction is $2.4 \mathrm{MHz}$. 
The modes along $y$ and $z$ direction do not affect the MS gate implementation considered
in the main text due to the following reasons. First, the axial modes ($z$-direction) do not couple
to the laser beam perpendicular to the ion chain. Second, the excitation of the radial modes in $y$-direction 
during the MS gate operation can be neglected providing that the secular frequency $\omega_y$
differs significantly (by hundreds of kilohertz) from $\omega_x$.

\section{The calculation of the trace over phonon modes}
\label{appendix:trace_calculation}

Here we present the necessary steps for the calculation of  $I(t,t')$ 
defined in Eq.~\eqref{eq:sensitivity_function_time}. By substituting the 
definition of $S^I_z(t)$ into Eq.~\eqref{eq:sensitivity_function_time},
one gets the expression
\begin{equation}
  I = I_1 - I_2,
  \label{eq:influence_function_verbose}
\end{equation}
where 
\begin{equation}
  I_1\left( t,t' \right) = \Tr_{\text{ph}}
  \left[ \langle \psi_{0}| 
  \hat U_0(t,t_0)^\dagger S_{z}\hat{U}_0(t,t')S_{z}
  \hat U_0(t', t_0)|\psi_{0} \rangle\right].
  \label{eq:I1}
\end{equation}
\begin{equation}
  I_2\left( t,t' \right) = \Tr_{\text{ph}}\left[
  \langle \psi_{0}|\hat U_0^\dagger(t,t_0) S_{z}U_0^\dagger(t,t_0)|\psi_{0} 
    \rangle
\langle \psi_{0}|\hat U_0^\dagger(t',t_0) S_{z}U_0(t',t_0)|\psi_{0} 
    \rangle
  \right].
  \label{eq:I2}
\end{equation}
Then, we substitute the evolution operator 
\eqref{eq:ms_evo_op} into \eqref{eq:influence_function_verbose}. To proceed, it is
necessary to express $S_z$ and $|\psi_0\rangle$ 
in the basis in the qubit subspace where all $\sigma_x^{(i)}$ are diagonal:
\begin{equation}
  S_z = \sum_{s} |s\rangle\langle s'| (S_z)_{ss'},
  \label{eq:sz_in_sx_basis}
\end{equation}
\begin{equation}
  |\psi_0\rangle = \sum_s (\psi_0)_s |s\rangle.
\end{equation}
Here $|s\rangle = |s_1s_2\rangle_x$ are the bit strings in the $x$-basis.

Decomposition in the $x$-basis allows using the analytical 
form of the evolution operator \eqref{eq:ms_evo_op}. The evolution operaror
can be expressed as a sum over qubit basis states as follows:
\begin{equation}
  \hat{U}_0 = \sum_s \hat{U}_s\otimes |s\rangle\langle s|,
  \label{eq:evo_op_decomposed}
\end{equation}
where
\begin{equation}
  \hat{U}_s(t_1, t_2) = 
    \exp\left(-\frac{i}{2}\sum_{i,j} \chi_{i,j}(t_2, t_1) 
    s_i s_j\right)\\
    \prod_m D_m\left(\sum_is_i\alpha_{im}(t_2, t_1)\right).
\end{equation}
Note that $\hat{U}_s(t_1,t_2)$ are the operators acting only on the phonon
part of the total Hilbert space. Also, the operator 
\eqref{eq:evo_op_decomposed} is diagonal in the qubit subspace.

After substituting \eqref{eq:sz_in_sx_basis} into \eqref{eq:I1} , 
one gets
\begin{equation}
I_1 = \sum_{s,s',s''}(\psi_0)^*_s (S_z)_{ss'} (S_z)_{s's''}
  (\psi_0)_{s''}
  \Tr_\mathrm{ph}\left[
    \hat{U}_s^\dagger(t,t_0) \hat{U}_{s'}(t,t')
    \hat{U}_{s''}(t',t'')
  \right],
  \label{eq:I1_expanded}
\end{equation}
\begin{equation}
I_2 = \sum_{s,s',s'', s'''}(\psi_0)^*_s (S_z)_{ss'} 
  (\psi_0)^*_{s'}(\psi_0)_{s''}
  (S_z)_{s''s'''}
  (\psi_0)_{s'''}
  \Tr_\mathrm{ph}\left[
    \hat{U}_s^\dagger(t,t_0) \hat{U}_{s'}(t,t_0)
    \hat{U}_{s''}^\dagger(t',t_0) \hat{U}_{s'''}(t',t_0)
  \right].
  \label{eq:I2_expanded}
\end{equation}
The trace of the products of the evolution operators reduces to the 
trace of products of multiple displacement operators, which can be done 
analytically \cite{Scully1997}. The number of terms 
in the summation over basis qubit states depends only on the number
of qubits in the Hamiltonian~\eqref{eq:ham_w_freq_noise} (two for
a two-qubit MS gate). Therefore, the
calculation of the expresssions 
\eqref{eq:I1_expanded} and \eqref{eq:I2_expanded} 
requires limited amount of the computational resources even for large 
number of ions in the chain.

\selectlanguage{russian}
\end{document}